\UseRawInputEncoding

\documentclass[reprint, twocolumn, showkeys]{revtex4}

\usepackage{graphicx}
\usepackage{dcolumn}
\usepackage{bm}
\usepackage{color}
\usepackage{amsmath,amsthm}
\usepackage{amssymb,bm}
\usepackage{mathrsfs}
\usepackage{comment}
\usepackage{ulem}

\graphicspath{%
    {converted_graphics/}
    {/}
}
\begin{document}

\title{Chiral cavity induced spin selectivity}
\author{Nguyen Thanh Phuc}
\email{nthanhphuc@moleng.kyoto-u.ac.jp}
\affiliation{Department of Molecular Engineering, Graduate School of Engineering, Kyoto University, Kyoto 615-8510, Japan}

\begin{abstract}
Chiral-induced spin selectivity (CISS) is a phenomenon in which electron spins are polarized as they are transported through chiral molecules, and the spin polarization depends on the handedness of the chiral molecule. 
In this study, we show that spin selectivity can be realized in achiral materials by coupling electrons to a single mode of a chiral optical cavity.
By investigating spin-dependent electron transport using the nonequilibrium Green's function approach, the spin polarization in a two-terminal setup is demonstrated to approach unity if the rate of dephasing is sufficiently small and the average chemical potential of the two leads is within an appropriate range of values, which is narrow because of the high frequency of the cavity mode.
To obtain a wider range of energies for a large spin polarization, we propose to combine the CISS in chiral molecules with the light-matter interactions.
For demonstration, the spin polarization of electrons transported through a helical molecule strongly coupled to a chiral cavity mode is evaluated. 
\end{abstract}


\maketitle

\textit{Introduction--}
CISS refers to the preferential transmission of electrons with one spin orientation over the other through chiral molecules~\cite{Naaman12, Naaman15, Naaman19}. 
Thus, chiral molecules can act as spin filters.
Since its discovery approximately two decades ago~\cite{Ray99}, CISS has been observed in numerous molecules and materials including DNA oligomers, oligopeptides, bacteriorhodopsin, helicenes and others~\cite{Xie11, Carmeli02, Gohler11, Mishra13, Kettner15, Kettner18, Lu19, Mishra19, Jia20}.
In addition to spintronic applications~\cite{Dor17, Suda19}, CISS can be used for enantiomer selection~\cite{Rosenberg15, Ghosh18} and to promote chemical reactions, such as electrochemical water splitting~\cite{Mtangi15, Mtangi17}.
However, using an Onsager-type argument, it can be shown that for electron transport through a single-orbital-per-site and two-terminal chiral system in the linear regime, CISS does not occur unless the time-reversal symmetry is broken~\cite{Yang19, Utsumi20, Evers22}.
This is known as the single-channel no-go theorem~\cite{Kiselev05, Bardarson08}.
Time-reversal symmetry can be effectively broken by non-unitary effects, such as dephasing or leakage~\cite{Guo12, Guo14, Matityahu13, Matityahu16}.
However, the effect of dephasing tends to destroy the quantum interference between different transport pathways, which plays a crucial role in CISS, as implicitly demonstrated by the fact that CISS cannot be observed in single-stranded DNA without electron's long-range tunneling~\cite{Gohler11, Mishra13}. 

On the other hand, it has recently been demonstrated that the physical and chemical properties of molecules and materials can be significantly modified by strongly coupling the system to an optical cavity~\cite{Ebbesen16, Ribeiro18, Feist18, Hertzog19, Herrera20, Nagarajan21, Li22, Basov21, Mivehvar21, Schlawin22}. 
The quantum fluctuation in the vacuum state of the cavity field offers a viable route for controlling matter in the dark, that is, without an external laser field.
Numerous intriguing phenomena arise from molecule-cavity coupling, including the manipulation of chemical landscapes in the excited-state manifold~\cite{Hutchison12, Herrera16, Galego16, Takahashi19}, modification of chemical reactivity by molecular-vibration polaritons~\cite{Thomas16, Hiura18, Thomas19, Lather19, Hirai20, Lather21, Galego19, Angulo19, Phuc20, Li20, Li21, Yang21}, superreaction~\cite{Phuc21}, Bose enhancement of energy transfer~\cite{Phuc22}, cavity-mediated superconductivity~\cite{Schlawin19}, and cavity-enhanced ferroelectricity~\cite{Ashida20, Latini21}. 

In this study, we show that CISS can be observed in achiral molecules and materials coupled to a single mode of a chiral optical cavity~\cite{Plum15, Mai19, Feis20, Taradin21, Hubener21, Gautier22, Voronin22}. 
Unlike the ordinary CISS in chiral molecules, under electron-cavity coupling, spin polarization can be nonzero without dephasing.
This is because the light-matter interaction can effectively break the time-reversal symmetry in the Hamiltonian of the electrons.
The nonequilibrium Green's function approach~\cite{Stefanucci-book, Jishi-book} is used to investigate spin-dependent electron transport in a two-terminal setup.
The spin polarization is close to unity if the average chemical potential of the two leads is within an appropriate range of values, which is narrow because of the high frequency of the cavity mode.
To obtain a wider range of energies for a large spin polarization, the CISS in chiral molecules can be combined with the light-matter interactions.
To demonstrate this, the spin polarization of the electrons transported through a helical molecule coupled to a chiral cavity mode is evaluated.
The dependence of the spin polarization on the length of the molecule, coupling strength, cavity frequency, and rate of dephasing is investigated.

\begin{figure}[tbp] 
  \centering
  \includegraphics[width=3in, keepaspectratio]{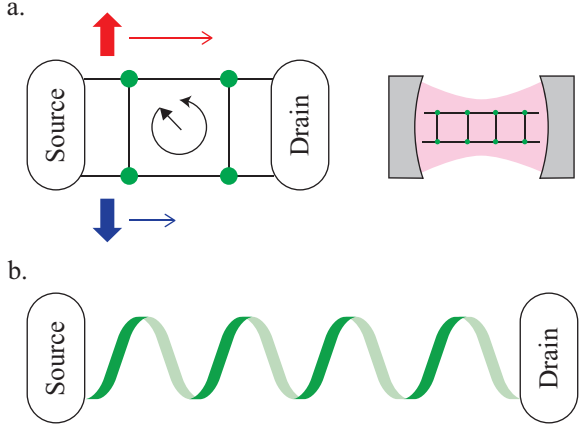}
  \caption{Illustration of the (a) chiral cavity induced spin selectivity and (b) ordinary CISS in chiral molecules. In (a), electrons being transported through an achiral system, such as a square 2D lattice, are coupled to a chiral optical cavity, whose vector potential rotates in the counter-clockwise direction. In both (a) and (b), the two-terminal setup in which the system is connected to a source and a drain is considered, and the spin polarization is reflected by the difference in the transmission of electrons with one spin orientation over the other.}
  \label{fig: system}
\end{figure}

\textit{Chiral cavity induced spin selectivity--}
We consider spin-dependent electron transport through a two-dimensional (2D) square lattice (of size $N_x\times N_y$ in the $xy$ plane) in a two-terminal setup, as illustrated in Fig.~\ref{fig: system}a. 
The Rashba spin-orbit coupling (SOC) with the potential gradient pointing along the $z$ direction has the following form: $\hat{H}_\text{SO}=\alpha(\hat{\sigma}_x\hat{p}_y-\hat{\sigma}_y\hat{p}_x)$, where $\alpha=-(e\hbar/4m_\text{e}^2c^2)(\text{d}V/\text{d}z)$, and $\hat{\boldsymbol{\sigma}}$ and $\hat{\mathbf{p}}$ are the spin Pauli matrices and momentum operators of electrons, respectively.
Electrons are coupled to a single mode of a chiral optical cavity, whose vector potential is given by $\hat{A}_x=A_0(\hat{a}+\hat{a}^\dagger)/\sqrt{2}$ and $\hat{A}_y=iA_0(\hat{a}-\hat{a}^\dagger)/\sqrt{2}$. 
Here, $A_0$ is the amplitude of the vector potential, and the operator $\hat{a}$ annihilates a photon in the cavity mode.
The interaction between electrons and the cavity field is considered in the lattice model using Peierls substitution~\cite{Peierls33}, where the tunneling amplitude between two sites centered at $\mathbf{r}_i$ and $\mathbf{r}_j$ is dressed by a factor $\exp\left\{(ie/\hbar)\int_{\mathbf{r}_i}^{\mathbf{r}_j}\hat{\mathbf{A}}\cdot\text{d}\mathbf{r}\right\}$.
In this study, we do not consider the ultrastrong coupling regime in which the optical diamagnetic term proportional to $\hat{\mathbf{A}}^2$ cannot be ignored~\cite{Kockum19, Phuc20}.
The total Hamiltonian of the coupled electron-cavity system is then given by $\hat{H}=\hat{H}_\text{ph}+\hat{H}_\text{e-ph}$, where $\hat{H}_\text{ph}=\hbar\omega_c\hat{a}^\dagger\hat{a}$ is the Hamiltonian for cavity photons with frequency $\omega_\text{c}$ and 
\begin{align}
\hat{H}_\text{e-ph}=&\sum_{j_x=1}^{N_x-1} \sum_{j_y=1}^{N_y}\sum_{s,s'}\Big\{\hat{c}_{j_x+1,j_y,s}^\dagger
\left[t_x\delta_{ss'}+i t_x^\text{SO}(\sigma_y)_{ss'}\right] \nonumber\\
&\times \hat{c}_{j_x,j_y,s'} e^{-ig_x(\hat{a}+\hat{a}^\dagger)} +\text{h.c.}\Big\} \nonumber\\
&+\sum_{j_x=1}^{N_x} \sum_{j_y=1}^{N_y-1}\sum_{s,s'}\Big\{\hat{c}_{j_x,j_y+1,s}^\dagger
\left[t_y\delta_{ss'}-i t_y^\text{SO}(\sigma_x)_{ss'}\right] \nonumber\\
&\times \hat{c}_{j_x,j_y,s'} e^{g_y(\hat{a}-\hat{a}^\dagger)} +\text{h.c.}\Big\}.
\end{align}
Here, $t_{x,y}$ and $t_{x,y}^\text{SO}$ are the spin-conserved and SOC-induced tunneling amplitudes, respectively; $g_{x,y}=eA_0d_{x,y}/\hbar\sqrt{2}$ are the electron-cavity coupling strengths in the $x$ and $y$ directions ($d_{x,y}$ are the lattice constants); h.c. stands for Hermitian conjugate; the site energy is set to zero without loss of generality.

In the two-terminal setup, the steady-state current of electrons with spin $s=\uparrow,\downarrow$ can be expressed in terms of nonequilibrium Green's functions as
\begin{align}
I_s=&\frac{ie}{2h}\int_{-\infty}^\infty \text{d}\omega \,\text{Tr}_s
\Big\{ \left[f_\text{L}(\omega)\Gamma^\text{L}(\omega)-f_\text{R}(\omega)\Gamma^\text{R}(\omega)\right] \nonumber\\
&\times \left[G^\text{R}(\omega)-G^\text{A}(\omega)\right]
+\left[\Gamma^\text{L}(\omega)-\Gamma^\text{R}(\omega)\right]G^<(\omega)\Big\},
\end{align}
where $G^{\text{R},\text{A},<}$ are the retarded, advanced, and lesser Green's functions of electrons, respectively; $\Gamma^{\text{L},\text{R}}$ represent the broadening matrices characterizing the coupling of electrons to the left and right leads; $f_{\text{L},\text{R}}$ are the Fermi-Dirac distribution functions of electrons in the two leads; $\text{Tr}_s$ denotes the trace over all electron states with spin $s$.
In the wide-band limit where the $\omega$-dependence is ignored, $\Gamma^\text{L}_{j_x,j_y,s;j_x',j_y',s'}=\Gamma_L\delta_{ss'}\delta_{j_x,1}\delta_{j_x',1}$ and $\Gamma^\text{R}_{j_x,j_y,s;j_x',j_y',s'}=\Gamma_R\delta_{ss'}\delta_{j_x,N_x}\delta_{j_x',N_x}$.
By applying Langreth's rule to the Dyson equation for the contour Green's function $G=G^0+G^0\Sigma G$, where $G^0$ is the Green's function without coupling to the leads, and $\Sigma$ is the proper self-energy, we obtain the Dyson equation for the retarded Green's function $G^\text{R}(\omega)=G^{0\text{R}}(\omega)+G^{0\text{R}}(\omega)\Sigma^\text{R}(\omega)G^\text{R}(\omega)$~\cite{Jishi-book}.
If the real part of the retarded self-energy, which only contributes to a shift in the electron energy, and the $\omega$-dependence are ignored, the retarded self-energy is related to the broadening matrices by $\Sigma^\text{R}=(-i/2\hbar)\left(\Gamma^\text{L}+\Gamma^\text{R}\right)$.
The Green's function $G^{0\text{R}}(\omega)$ can be obtained by diagonalizing the Hamiltonian of the coupled electron-cavity system $\hat{U}^{-1}\hat{H}\hat{U}=\text{diag}(\epsilon_1,\cdots,\epsilon_N)$, where $N$ is the dimension of the hybrid system with a truncated photon number.
Outside the ultrastrong coupling regime, we assume that the modification of the ground state of the hybrid system by the light-matter interaction is negligible, and the cavity field is in the vacuum state $|0\rangle_\text{p}$.
By denoting $\alpha=|j_x,j_y,s\rangle\otimes |0\rangle_\text{p}$ and $\beta=|j_x',j_y',s'\rangle\otimes |0\rangle_\text{p}$ as the tensor products of an electronic and photonic state, the matrix elements of $G^{0\text{R}}(\omega)$ are given by $G^{0\text{R}}_{j_x,j_y,s;j_x',j_y',s'}(\omega)=\sum_{\gamma=1}^N U_{\alpha\gamma}U_{\beta\gamma}^*/(\omega-\epsilon_\gamma/\hbar+i\eta)$, where $\eta$ is an infinitesimal positive number.
Similarly, the Dyson equation for the lesser Green's function of noninteracting electrons is given by $G^<(\omega)=G^\text{R}(\omega)\Sigma^<(\omega)G^\text{A}(\omega)$, where $\Sigma^<(\omega)=(i/\hbar)\left[f_\text{L}(\omega)\Gamma^\text{L}+f_\text{R}(\omega)\Gamma^\text{R}\right]$ and $G^\text{A}(\omega)=[G^\text{R}(\omega)]^\dagger$. 
If a bias voltage $V_\text{b}$ is applied, the chemical potentials of the two leads are given by $\mu_\text{L,R}=\bar{\mu}\pm eV_\text{b}/2$ where $\bar{\mu}$ is the average chemical potential.
The differential conductance $c_s\equiv(\text{d}I_s/\text{d}V_\text{b})_{V_\text{b}=0}$ at zero temperature, for which the Fermi-Dirac distribution function reduces to the step function, is then given by
\begin{align}
c_s=&\frac{ie^2}{4h}\Big[ \text{Tr}_s\left\{\left(\Gamma^L+\Gamma^R\right)\left[G^\text{R}(\bar{\mu})-G^\text{A}(\bar{\mu})\right]\right\}\nonumber\\
&+\frac{i}{\hbar} \text{Tr}_s\left\{\left(\Gamma^\text{L}-\Gamma^\text{R}\right)G^\text{R}(\bar{\mu})\left(\Gamma^\text{L}-\Gamma^\text{R}\right)G^\text{R}(\bar{\mu})\right\}\Big].
\end{align}
The spin polarization is defined by $P_\text{s}=(c_\uparrow-c_\downarrow)/(c_\uparrow+c_\downarrow)$.
In the following calculations, the system parameters are set to $t_x=t_y=0.1\,\text{eV}$, $t_x^\text{SO}=t_y^\text{SO}=0.012\,\text{eV}$, and $\Gamma_\text{L}=\Gamma_\text{R}=1\,\text{eV}$, which are typical orders of magnitude for electron transport in organic molecules, such as DNA and proteins~\cite{Yan02, Endres04, Senthilkumar05, Hawke10}.
The finite lifetime of photon is irrelevant for the cavity field in the vacuum state.

\begin{figure}[tbp] 
  \centering
  \includegraphics[width=3.4in, keepaspectratio]{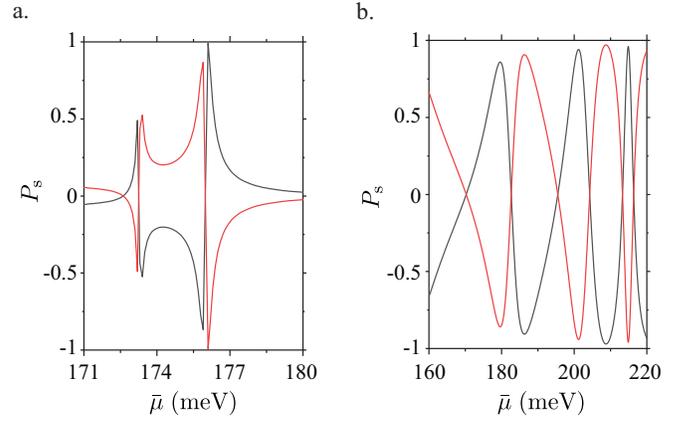}
  \caption{Spin polarization $P_\text{s}$ of electrons in a two-terminal transport through (a) a square lattice and (b) a helical molecule as a function of the average chemical potential $\bar{\mu}$ of the two leads. In both cases, the electrons are coupled to a chiral cavity mode with frequency $\omega_\text{c}=1\,\text{eV}/\hbar$ and the coupling strength is $g=0.1\,\text{eV}$. The red and black curves correspond to the cavity modes circularly polarized in the clockwise and anticlockwise directions, respectively. Here, the chemical potential is measured relative to the site energy of electrons.}
  \label{fig: spin polarization for square lattice and helical molecule}
\end{figure}

Figure~\ref{fig: spin polarization for square lattice and helical molecule}a shows the spin polarization as a function of the average chemical potential in the range $171\,\text{meV}<\bar{\mu}<180\,\text{meV}$ for $N_x=30$ and $N_y=2$.
The cavity frequency and coupling strength were set to $\omega_\text{c}=1\,\text{eV}/\hbar$ and $g_x=g_y=0.1\,\text{eV}$, respectively, and the dephasing was ignored.
It is clear that, unlike the ordinary CISS in chiral molecules, the spin polarization is nonzero without dephasing.
This is because the time-reversal symmetry is effectively broken in the electron system by the light-matter interaction.
This can be understood by considering the high-frequency limit, where $\omega_\text{c}$ is larger than the other system parameters.
In this limit, Van Vleck degenerate perturbation theory can be used to derive the effective Hamiltonian for electrons~\cite{Shavitt80, Eckardt15}.
Up to the first order in $1/\omega_\text{c}$, the spin-dependent effective tunneling amplitude between the $i$th and $j$th sites is given by
\begin{align}
t^\text{eff}_{i,s;j,s'}=&t_{i,s;j,s'} \,_\text{p}\langle 0|\hat{D}_{ij}|0\rangle_\text{p}
-\sum_{n\not=0}\sum_{k\not=i,j}\sum_{s"} \frac{t_{i,s;k,s"}t_{k,s",j,s'}}{n\hbar\omega_\text{c}} \nonumber\\
&\times \,_\text{p}\langle 0|\hat{D}_{ik}|n\rangle_\text{p} \,_\text{p}\langle n|\hat{D}_{kj}|0\rangle_\text{p},
\label{eq: effective tunneling amplitude}
\end{align}
where $t_{i,s;j,s'}$ is the tunneling amplitude in the absence of cavity coupling and the associated photonic displacement operator is given by $\hat{D}_{ij}=e^{-ig_x(\hat{a}+\hat{a}^\dagger)}$ and $\hat{D}_{ij}=e^{g_y(\hat{a}-\hat{a}^\dagger)}$ for tunneling in the $x$ and $y$ directions, respectively. 
The matrix elements of the displacement operator $\hat{D}(\alpha)=e^{\alpha\hat{a}^\dagger-\alpha^*\hat{a}}$ are calculated as
\begin{align}
&\langle n|\hat{D}(\alpha)|m\rangle\nonumber\\
&=\left\{
\begin{array}{rcl}
\sqrt{\frac{m!}{n!}}e^{-|\alpha|^2/2}\alpha^{n-m}L_m^{n-m}(|\alpha|^2) &\mbox{for} & n\geq m \\
\sqrt{\frac{n!}{m!}}e^{-|\alpha|^2/2}(-\alpha^*)^{m-n}L_n^{m-n}(|\alpha|^2) &\mbox{for} & m\geq n
\end{array}
\right.,
\label{eq: matrix element of displacement operator}
\end{align}
where $L_n^m(x)=\sum_{k=0}^n (-1)^k \left( \begin{array}{cc} n+m\\k+m \end{array} \right)x^k/k!$ denotes the associated Laguerre polynomial.
It is evident from Eqs.~\eqref{eq: effective tunneling amplitude} and ~\eqref{eq: matrix element of displacement operator} that a nonzero phase emerges in the effective tunneling of electrons in one closed loop, thereby breaking the time-reversal symmetry.
As shown in Fig.~\ref{fig: spin polarization for square lattice and helical molecule}a, $P_\text{s}$ changes its sign when the circular polarization of the cavity mode changes its direction, indicating that spin polarization stems from the chirality of the cavity mode. 
Moreover, $P_\text{s}$ can be close to unity if $\bar{\mu}$ is within an appropriate range of values; however, it is narrow compared to the ordinary CISS in chiral molecules~\cite{Guo12, Guo14} because of the high frequency of the cavity mode.

To obtain a wider energy range for large spin polarization, the static chirality of molecules can be combined with the dynamic chirality of light.
As an example, the CISS in a helical molecule coupled to a chiral cavity mode is considered as follows.
Electron transport through a helical molecule is described by a 1D Hubbard model with a spatially dependent SOC-induced tunneling amplitude~\cite{Guo14}.
The Hamiltonian is given by
\begin{align}
\hat{H}_\text{SO}=\sum_{j=1}^{N-1}\sum_{k=1}^{N-j}\sum_{s,s'}
\left(2i t^\text{SO}_k\cos\varphi_{j,k}^- \hat{c}^\dagger_{j+k,s}\sigma^{j,k}_{ss'}\hat{c}_{j,s'} +\text{h.c.}\right),
\end{align}
where $N$ is the number of sites, $t^\text{SO}_k$ is the tunneling amplitude over $k$ sites, $\sigma^{j,k}=\left(\sigma_x\sin\varphi_{j,k}^+-\sigma_y\cos\varphi_{j,k}^+\right)\sin\theta_k+\sigma_z\cos\theta_k$ with $\varphi_{j,k}^\pm=(\varphi_{j+k}\pm\varphi_j)/2$ and $\varphi_j=j\Delta\varphi$. 
Here, $\Delta\varphi$ is the twist angle between the nearest neighbor sites, and $\theta_k$ is the angle between the vector connecting $k$-neighbor sites and the $xy$ plane orthogonal to the helical axis ($z$ axis).
For CISS to occur in a single helix, long-range tunneling of electrons is required, implying the importance of quantum interference between different transport pathways~\cite{Gohler11, Mishra13}.
The tunneling amplitude is assumed to decay exponentially with distance.
For comparison, the values of the system parameters were chosen to be the same as those for the $\alpha$-helical protein studied in Ref.~\cite{Guo14}.
The electron-cavity coupling is considered using Peierls substitution in the same way as above.
Figure~\ref{fig: spin polarization for square lattice and helical molecule}b shows the spin polarization as a function of the average chemical potential for $N=30$.
The values of the cavity frequency and coupling strength $g=eA_0R\sqrt{2}/\hbar$, where $R$ denotes the helix radius, are the same as those in the case of a square lattice.
A wider range of energies is observed for a large spin polarization than in the case of a square lattice.

\begin{figure}[tbp] 
  \centering
  \includegraphics[width=3.4in, keepaspectratio]{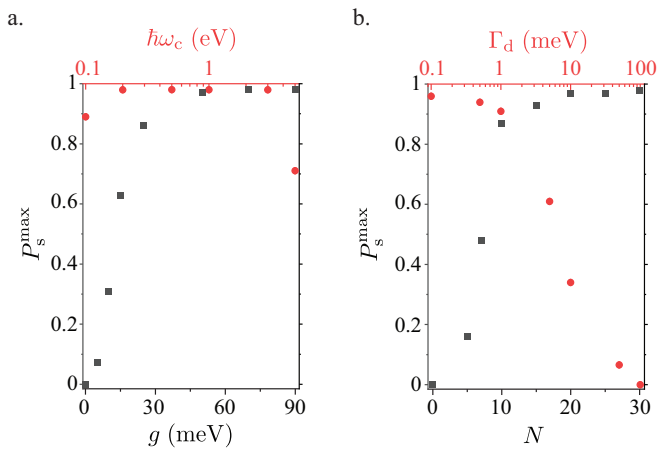}
  \caption{Dependence of the maximum spin polarization $P_\text{s}^\text{max}$ of electron transport through a helical molecule coupled to a chiral optical cavity on (a) coupling strength $g$, cavity frequency $\omega_\text{c}$, (b) length of the molecule (the number of atomic sites $N$), and rate of dephasing $\Gamma_\text{d}$.}
  \label{fig: dependences of the maximum spin polarization on various parameters}
\end{figure}

The dependence of the maximum spin polarization (over the variable average chemical potential) on the coupling strength and cavity frequency is shown in Fig.~\ref{fig: dependences of the maximum spin polarization on various parameters}a, from which it is evident that the spin polarization increases with the coupling strength, and $P_\text{s}^\text{max}$ is close to unity at a moderate coupling strength $g\simeq 50\,\text{meV}$.
In contrast, the maximum spin polarization is almost independent of the cavity frequency in a wide range $0.2\,\text{eV}\leq \hbar\omega_\text{c} \leq 3\,\text{eV}$.
The dependence of $P_\text{s}^\text{max}$ on the length of the molecule ($N$ atomic sites) is shown in Fig.~\ref{fig: dependences of the maximum spin polarization on various parameters}b.
It is clear that the maximum spin polarization increases with $N$ and $P_\text{s}^\text{max}$ can be large, even for a relatively short molecule with $N=10$.
Furthermore, in the presence of cavity coupling, electrons can be transported through the system when the average chemical potential of the leads is higher than the site energy of the system by an integer multiple of $\hbar\omega_\text{c}$.
Physically, the excess energy of the electrons in the leads can be used to create photons in the cavity.
However, electrons cannot be transported through the system if the chemical potential is lower than the site energy of the system. 
This is because photon absorption cannot occur when the cavity field is in the vacuum state. 

Finally, the dependence of the spin polarization on the dephasing rate is investigated. 
Dephasing stems from the inelastic scattering of electrons with lattice or molecular vibrations, namely phonons, and results in the loss of phase and spin memory of the electrons.
To simulate the effect of dephasing, Buttiker's virtual leads are introduced at each site with broadening $\Gamma_\text{d}$~\cite{Datta-book}.
Because the net currents through the virtual leads are zero, the voltages of virtual leads can be calculated using the Landauer-Buttiker formula for the current in the $q$th lead (real or virtual) with spin $s$ at energy $\epsilon$: $I_{q,s}(\epsilon)=(e^2/h)\sum_m T_{q,m,s}(\epsilon)(V_m-V_q)$, where $V_q$ is the voltage in the $q$th lead and $T_{q,m,s}(\epsilon)=\text{Tr}_s\left\{ \Gamma^q G^\text{R}(\epsilon)\Gamma^m G^\text{A}(\epsilon)\right\}$ is the transmission coefficient from the $m$th to the $q$th lead.
Here, $\Gamma^q$ is the broadening matrix associated with the $q$th lead.
The electronic Green's functions can be obtained from the effective Hamiltonian for electrons in the high-frequency limit using $G^\text{R}(\epsilon)=[G^\text{A}(\epsilon)]^\dagger=\left[\epsilon I-\hat{H}_\text{eff}-\sum_{q}\Sigma^\text{R}_q\right]^{-1}$ with $\Sigma^\text{R}_q=-i\Gamma^q/2$.
Figure~\ref{fig: dependences of the maximum spin polarization on various parameters}b shows the maximum spin polarization as a function of $\Gamma_\text{d}$, which is proportional to the rate of dephasing.
Evidently, $P_\text{s}^\text{max}$ increases monotonically with a decreasing rate of dephasing, in contrast to ordinary CISS without cavity coupling.
Therefore, the spin polarization can systematically be increased, for example, by lowering the temperature.
For electron transport through a molecule at room temperature, wherer the rate of dephasing is $\Gamma_\text{d}\simeq 5\,\text{meV}$~\cite{Guo12, Guo14}, $P_\text{s}^\text{max}$ can be as high as 0.6.

\textit{Conclusion--}
We have shown that CISS in the two-terminal electron transport can be realized with achiral molecules and materials if the system is coupled to a chiral optical cavity and the average chemical potential of the two leads is within an appropriate range of values, which is narrow because of the high frequency of the cavity.
Unlike ordinary CISS in chiral molecules, spin polarization is nonzero without dephasing because the light-matter interaction can effectively break the time-reversal symmetry in the Hamiltonian of the electrons.
To obtain a wider range of energies for large spin polarization, chiral molecules can be combined with light-matter interactions. 
The spin polarization can approach unity for sufficiently strong electron-cavity coupling and a low rate of dephasing, which are well within the reach of current experiments.
Chiral cavity induced spin selectivity offers a powerful tool for controlling spin dynamics in various molecules and materials with potential applications in spintronics and chemical reactions.
Finally, the optical control of the CISS can be realized alternatively with an external laser field in the context of Floquet engineering~\cite{Phuc18, Phuc19}.


\begin{acknowledgements}
The computations were performed using Research Center for Computational Science, Okazaki, Japan.
\end{acknowledgements}



\end{document}